# Evidence of Flat Bands and Correlated States

# in Buckled Graphene Superlattices

Jinhai Mao[1,5], Slaviša P. Milovanović[2], Miša Anđelković[2], Xinyuan Lai[1], Yang Cao[3], Kenji Watanabe[4], Takashi Taniguchi[4], Lucian Covaci[2], Francois M. Peeters[2], Andre K. Geim[3], Yuhang Jiang[1,6*], and Eva Y. Andrei[1*]

[1]Department of Physics and Astronomy, Rutgers University, 136 Frelinghuysen Road, Piscataway, NJ 08855 USA
[2]Departement Fysica, Universiteit Antwerpen, Groenenborgerlaan 171, B-2020 Antwerpen, Belgium
[3]School of Physics and Astronomy, University of Manchester, Oxford Road, Manchester M13 9PL, United Kingdom
[4]Advanced Materials Laboratory, National Institute for Materials Science, 1-1 Namiki, Tsukuba 305-0044, Japan
[5]School of Physical Sciences and CAS Center for Excellence in Topological Quantum Computation, University of Chinese Academy of Sciences, Beijing, China
[6]College of Materials Science and Opto-Electronic Technology, University of Chinese Academy of Sciences, Beijing, China

**Two-dimensional (2D) atomic crystals can radically change their properties in response to external influences such as substrate orientation or strain, resulting in essentially new materials in terms of the electronic structure[1-5]. A striking example is the creation of flat-bands in bilayer-graphene for certain "magic" twist-angles between the orientations of the two layers[6]. The quenched kinetic-energy in these flat-bands promotes electron-electron interactions and facilitates the emergence of strongly-correlated phases such as superconductivity and correlated-insulators. However, the exquisite fine-tuning required for finding the magic-angle where flat-bands appear in twisted-bilayer graphene, poses challenges to fabrication and scalability. Here we present an alternative route to creating flat-bands that does not involve fine tuning. Using scanning tunneling microscopy and spectroscopy, together with numerical simulations, we demonstrate that graphene monolayers placed on an atomically-flat substrate can be forced to undergo a buckling-transition[7-9], resulting in a periodically modulated pseudo-magnetic field[10-14], which in turn creates a post-graphene material with flat electronic bands. Bringing the Fermi-level into these flat-bands by electrostatic doping, we observe a pseudogap-like depletion in the density-of-states (DOS), which signals the emergence of a correlated-state[15-17]. The described approach of 2D crystal buckling offers a strategy for creating other superlattice systems and, in particular, for exploring interaction phenomena characteristic of flat-bands.**

Weakly dispersive, "flat" bands, facilitate the emergence of strongly correlated electronic phases that could survive up to high temperatures[18-20]. A celebrated example is the Landau-level sequence of magnetically-induced flat-bands that can host correlated phases such as fractional quantum-Hall states[21,22] or magnetically-induced Wigner crystals[23]. Magnetically-induced flat-bands however, have limited applicability because the broken time-reversal symmetry precludes the emergence of certain correlated-states such as superconductivity. More recently, twisted bilayer graphene that is finely-tuned to a "magic angle" where a flat-band emerges, has introduced a new platform for the creation of correlated phases[4,5].

Here we explore an alternative path to flat bands that does not require fine-tuning or breaking time-reversal symmetry. The strategy involves creating flat bands by utilizing the band-structure reconstruction that occurs when a 2D membrane undergoes a buckling transition[14 19,20]. Buckling transitions in stiff membranes are typically triggered by in-plane compressive strain[7-9] that can be generated during thermal cycling, by solvent-induced capillary forces, or by substrate-induced stress. Upon exceeding a critical strain value, buckling of the membrane reduces its elastic energy through out of plane distortions resulting in intriguing periodic strain patterns (Figure 1b, c) whose structure is dictated by boundary conditions and strain distribution[9]. We find that in graphene, buckling-induced strain arrays gives rise to a periodic pseudo-magnetic field (PMF)[10-13] which reconstructs the low-energy band structure into a series of essentially flat-bands. Unlike earlier realizations of PMF which were mostly local in nature[24], the buckling-transitions studied here produce a global change in the electronic-structure comprised of a sequence of flat-bands that percolate throughout the material.

We employed the thermally induced buckling-transition in graphene deposited on $NbSe_2$ or hBN substrates to create a periodic PMF. The buckling structures studied here are typically nested between ridges (Extended Data Figure 1) that often form in graphene during fabrication. Topographical analysis suggests that the buckling is triggered by the compressive strain generated by the collapse of ridges during thermal cycling (Sample Preparation). This produces various buckling-patterns, from 1D to 2D, with similar nanometer-scale periods (Figure 1b, c), suggesting a universal buckling-mechanism that is insensitive to the lattice mismatch between graphene and its substrate.

In Figure 1a we show the schematics of the sample and scanning tunneling microscopy (STM) measurement setup. The topography of a buckling-induced triangular super-structure in graphene deposited on $NbSe_2$ is shown in Figure 1b (right panel). This superlattice consists of alternating crests (bright) and troughs (dark) (Figure 2a) with 0.17 nm height modulation and period $a_B = 14.4 \pm 0.5\, nm$. The large lattice mismatch between graphene (0.246 nm) and NbSe$_2$ (0.36 nm) rules out a moiré pattern[2] interpretation (Methods). We first focus on the electronic structure obtained from the *dI/dV* spectra (*I* is the current, *V* is the bias) in the crest regions (Figure 2b). The spectra comprise a sequence of peaks identified as PMF-induced pseudo Landau levels (PLLs)[14,24,25]. The energy of the most pronounced peak (labeled *N*=0) is aligned with the charge neutrality point (CNP) as determined from the spectra taken in the un-buckled area (Extended Data Figure 2A). The CNP energy, $E_D$, is shifted by ~ 0.5 eV above the Fermi level, due to the hole doping induced by the NbSe$_2$ substrate[26]. Labeling the remaining PLL peaks in order of increasing (decreasing) energies as *N* = ±1, ±2, ... we find that they follow the sequence expected of Landau levels in flat monolayer graphene[27], $E_N = E_D + sgn(N)v_F\sqrt{2e\hbar|N|B_{PMF}}$. This PLL sequence confirms the existence of a PMF in the buckled graphene membrane. Using the Fermi-velocity value, $v_F = 1.0 \times 10^6\, m/s$, we estimate the PMF in the center of the crest region as $B_{PMF} = 108 \pm 8$ T (Figure 2b, Right inset). Outside the buckled region where the STM topography is flat, the spectrum exhibits a featureless V-shape as expected of unstrained graphene on NbSe$_2$ with the minimum at E$_D$ ~ 0.5 eV (Extended Data Figure 2).

Another hallmark of a PMF is the sublattice polarization of the electronic wavefunction in the *N*=0 level, where the two sublattices host opposite signs of the PMF. This means that for a given PMF sign, the *N*=0 electronic wavefunction is localized on one sublattice, while for the opposite sign PMF it resides on the other sublattice[12,28,29]. The PMF-induced sublattice polarization, which is the counterpart of the magnetically-induced valley polarization of the *N*=0 level, follows directly from the opposite signs of the PMF in the K and K' valleys. The observation of sublattice polarization in the atomic-resolution STM images shown in Figures 2d-f, thus provides a direct experimental signature of the extant PMF in buckled graphene superstructures [28-32].

In Figures 2d-f we show atomic-resolution STM topography in the crest, trough and intermediate regions of the superlattice. In both crests (Figure 2e) and troughs (Figure 2f), the sublattice polarization is clearly revealed by the triangular structure reflecting the fact that only atoms in one sublattice, say A, are visible in the crests while only the atoms in the B sublattice are visible in the troughs. This in turn confirms that the sign of the PMF in the crests is opposite to that in the troughs. In the transition regions between crests and the troughs (Figure 2d), we observe a honeycomb pattern where both A and B sublattices are visible, indicating a zero-crossing of the PMF. This PMF sublattice polarization is closely reproduced in the numerical simulations (Figure 3) discussed below.

Turning to the *dI/dV* spectra in the trough regions (Figure 2c), we note that they also exhibit a sequence of peaks. Plotting the energy of these peaks versus the level-index we find that the sequence is essentially linear in *N*, with a roughly equidistant energy spacing of ~ 89 ± 2 meV.

To better understand these results, we performed tight-binding calculations in the presence of a periodic PMF with a triangular structure similar to that in the experiment (Figure 3d):

$$\boldsymbol{B}_{PMF}(x,y) = B[\cos(\boldsymbol{b_1 r}) + \cos(\boldsymbol{b_2 r}) + \cos(\boldsymbol{b_3 r})] \tag{1}$$

Here *B* is the PMF amplitude, $\boldsymbol{b_1} = \frac{2\pi}{a_b}\left(1, -\frac{1}{\sqrt{3}}\right)$, $\boldsymbol{b_2} = \frac{2\pi}{a_b}\left(0, \frac{2}{\sqrt{3}}\right)$, $\boldsymbol{b_3} = \boldsymbol{b_1} + \boldsymbol{b_2}$, and $a_b = 14$ nm. This choice of coordinates reflects the C6 symmetry of the PMF superlattice and the fact that the total PMF flux has to vanish, because time-reversal symmetry is not broken. It corresponds to an array of PMF crests peaked at a maximum PMF of $3B$, surrounded by a percolating network of troughs where the minimum PMF is $-1.5B$. The zeros of this PMF configuration form circles surrounding each crest. Because of the spatial PMF variation, the measured spectra correspond to an effective PMF value, $B_{eff}$, which approximately averages the field over the pseudo-cyclotron orbit. This averaging effect becomes more pronounced as the ratio between the magnetic-length and the lattice-period decreases (Extended Data Figure 6). We found that the PLL sequence obtained for $B_{eff} = 112$ T ($B = 120$ T) matches the experimentally measured sequence shown in Figure 2b.

Figures 3a-c plot the evolution of the calculated local density of states (LDOS) with PMF amplitude, for each sublattice in the crest, transition and trough regions. Interestingly, in the crest and trough regions we observe a strong imbalance in the low energy LDOS intensity between the A and B sublattices (Figure 3a, and c), which is absent in the intermediate transition region (Figure 3b). This is consistent with the experimentally observed sublattice polarization in the crest (Figure 2e) and trough regions (Figure 2f), and with its absence in the transition region (Figure 2d). In Figure 3e we show the simulated LDOS spectrum in the crest region for $B_{eff} = 112$ T together with a fit to the square-root dependence on *N*, which is consistent with the experimentally measured spectra shown in Figure 2b as discussed in Extended Data Figure 6. The simulated LDOS spectrum in the trough region (Figure 3f) is approximately linear in *N* with a level spacing of ~ 90 meV, consistent with the experimental results in Figure 2c. To elucidate the origin of this linear peak sequence, we show in Figure 3g the evolution of the LDOS calculated along a path connecting two crests (arrow in Figure 3d bottom). The experimental peak positions in the crest (trough) regions are shown by right-pointing (left-pointing) arrows respectively. In the center of the trough the equidistant level sequence is clearly seen. We note that these levels are not solely determined by the local value of the PMF and they do not exhibit spatial dispersion. Furthermore, although these states spread into the crest regions, they disappear upon approaching the PLLs in the crest center. The discrete nature of these equidistant levels indicates that they originate from strain-induced confinement within the quantum well defined by the PMF, closely resembling magnetic-confinement in quantum-dots in 2D semiconductors[33,34]. As in the case of quantum-dots, here the electrons are trapped in a PMF-induced potential-well, which produces a set of levels spaced by a characteristic (geometry-dependent) energy scale $\Delta E \sim \frac{\hbar v_F \pi}{W}$, where $W$ is the dot size. Using the energy-scale of the levels in the trough region, ~ 90 meV, we estimate $W \sim 21$ nm, which is approximately the size of the well indicated by the grey dashed lines in Figure 3g bottom. In the trough region, the energy of each level decreases with increasing PMF (Figure 3c) until, when the magnetic-length becomes considerably smaller than the dot size, the levels all merge into one degenerate level that approaches the CNP.

We next discuss the emergence of flat-bands in this system. The periodic potential imposed by the PMF superlattice breaks up the low energy conical band of graphene into a series of mini-bands whose width is controlled by the strength of the PMF amplitude, $B$. At low values of $B$, the minibands restructure the LDOS into a series of semi-discrete levels as illustrated in Extended Data Figure 7. As $B$ increases, these levels evolve into increasingly narrower bands that become flat in the limit of large $B$. In Figure 4a we plot the first few minibands in the buckled G/NbSe$_2$ sample for $B_{eff} = 112$ T. They all show flat-band segments along the K'-M' line in the mini-Brillouin zone. The corresponding simulated LDOS contours in the three minibands with energies $E_0$-$E_D$ = -0.03 eV, $E_1$-$E_D$ = -0.17 eV and $E_2$-$E_D$ = -0.28 eV, shown in Figure 4b, show good agreement with the measured $dI/dV$ maps in Figure 4c, indicating that the model captures the salient features of the data. This sequence of buckling induced flat-bands that are well separated from each other, would be ideally suited for hosting correlated electronic states, if it were possible to align the Fermi energy within one of the flat-bands. However, this was not possible in the buckled G/NbSe$_2$ sample, because of the finite conductance of the NbSe$_2$ substrate. Nevertheless, the observation of the buckling induced flat-bands in this sample indicates that this could be achieved by creating a buckling-pattern in graphene deposited on an insulating substrate, as shown below.

In order to study correlation effects in buckling-induced flat-bands, we turn to the G/hBN sample. In this case, applying a gate voltage across the insulating hBN layer allows us to bring the Fermi level within the flat-band. Utilizing the same sample preparation process as for the G/NbSe$_2$ sample, we again obtain buckling superlattices as illustrated in Figure 5a, b.

The gate-voltage dependence of the *dI/dV* spectra in the transition region where the strain is minimal (Figure 5c) shows that for V$_g$ =0 the spectrum is V-shaped and its minimum, which marks the CNP, is aligned with the Fermi level. This indicates that, unlike the case of G/NbSe$_2$, the G/hBN sample is not doped by the hBN substrate. Tuning the gate-voltage from -62 V to +38 V, gradually moves the CNP from the hole-doped (~ -100 mV) to the electron-doped sector (~ +200 mV). In Figure 5d we compare the *dI/dV* spectra in the crest and transition regions for the heavily hole-doped ($V_g = -62\,V$) case. Here, the crest spectrum features a prominent peak at the CNP that is flanked by a sequence of weaker peaks (blue arrows), closely resembling the crest spectrum in the G/NbSe$_2$ sample (Figure 2b). As before, we identify the prominent peak with the strain-induced *N*=0 PLL, and fitting the peak sequence with a square-root level-index dependence, we obtain the PMF value $B_{eff} \sim 104\,T$ (inset of Figure 5d) consistent with the simulations described in Extended Data Figure 9. Doping the sample from the hole-doped (V$_g$ = -62 V) to the electron-doped ($V_g = +38\,V$) sector (Figure 5e) we note that the *N*=0 PLL tracks the CNP. As before the *N*=0 PLL corresponds to a buckling induced weakly dispersive flat-band. When doping the sample to partially fill this band, a pseudo-gap feature at the Fermi level splits the peak in two, indicating the appearance of a correlation induced state. Labeling the peaks above and below the Fermi level as upper band (UB), and lower band (LB), we find that when the Fermi level is aligned with the charge neutrality point, the UB and LB peaks have equal intensities. Doping away from charge neutrality we observe a pronounced spectral weight redistribution between the two peaks, so that in the electron doped regime (V$_g$ > 0) the LB intensity dominates over the UB, whereas in the hole doped regime (V$_g$ < 0) the UB becomes dominant. The appearance of the pseudo-gap feature and the spectral weight redistribution in the partially filled flat-band in the crest regions of buckled G/hBN is strikingly similar to that observed in the partially filled flat-band of magic angle twisted bilayer graphene[15-17] where correlation induced insulating and superconducting states have been observed.

These findings demonstrate that buckling-induced periodic strain patterns offer a new experimental strategy for the creation of flat-bands and for inducing correlated states with exceptional flexibility. The shape, period and symmetry of the buckled structures can be controlled by experimentally adjustable parameters, such as boundary geometry and strain distribution, enabling the realization of flat-bands with prescribed geometry[35]. We believe that the described approach to buckle 2D crystals will be used widely for creating other superlattice systems with controllable electronic band structure for exploring the effects of strong interactions and the emergence of correlated phases.

**Acknowledgments**

We acknowledge support from DOE-FG02-99ER45742 (E.Y.A., Y.J.), Gordon and Betty Moore Foundation GBMF9453 (E.Y.A), National Key R&D Program of China (2019YFA0307800, 2018YFA0305800) (J.M.), Beijing Natural Science Foundation (Z190011) (J.M.), Flemish Science Foundation FWO-Vl (S.P.M and F.M.P), TRANS2DTMD Flag-Era project (M.A, L.C. and F.M.P). We thank Francisco Guinea, Benjamin Davidovitch and Dominic Vella for stimulating discussions during the Aspen winter workshop on Low-dimensional solids in hard and soft condensed matter.

**Author Contributions**

J. M. and Y. J. performed STM/STS measurements. Y. J., J. M. and E. Y. A. designed the research strategy, performed data analysis and wrote the manuscript with input from all authors. S. P. M., M. A., L.C. and F. M. P performed theoretical calculations. Y. C., A. K. G and X. L. fabricated the devices. K. W. and T. T. provided hexagonal boron nitride. E. Y. A. supervised the project.

**Author Information**

Reprints and permissions information is available at www.nature.com/reprints. The authors declare no competing interests. Readers are welcome to comment on the online version of the paper. Publisher's note: Springer Nature remains neutral with regard to jurisdictional claims in published maps and institutional affiliations. Correspondence and requests for materials should be addressed to Y. J. (yuhangjiang@ucas.edu.cn) and E. Y. A. (andrei@physics.rutgers.edu).

**Main Figure Legends:**

**Figure 1. Buckled structures in graphene membranes. a,** Schematics of sample geometry and measurement setup. **b,** Buckling-modes observed by STM topography in G/NbSe$_2$ sample. Right: Topography of G/NbSe$_2$ showing a triangular buckling pattern ($V_b$ = 0.5 V, $I$ = 40 pA). Inset: Height profile along the arrow shows the ~0.17 nm height modulation of the buckling pattern with lateral period 14.4 nm $\pm$0.5 nm. **c**, 1D and 2D buckling modes observed by STM topography in G/hBN sample ($V_b$ = -300 mV, $I$ = 20 pA). Grey dashed lines are guides to eye marking the periodic structure.

**Figure 2. Pseudo Landau level quantization and sublattice polarization in buckled graphene. a,** STM topography (70 nm×70 nm) of a buckled graphene membrane supported on NbSe$_2$ reveals a 2D triangular lattice ($V_b$ = 0. 6V, $I$ = 20 pA). **b,** $dI/dV$ spectrum taken in the crest region labeled by the red square in the left inset. Right inset: PLL energy plotted against the square-root of the LL index, $N = 0, \pm1, \pm2, ....$ **c**, Same as **b** taken in a trough region (blue square in the left inset of **b**). The negative second derivative of the $dI/dV$ signal (black curve) is superposed to better reveal the peak energies (black arrows). Inset shows the linear dependence of the peak energy on the level index, $N$. **d-f**, Atomic resolution STM topography in transition (orange rectangle), crest (red square in left inset of **b**) and trough (blue square) regions (color scale shown in panel **f**). A schematic honeycomb lattice with the yellow and green balls representing the two sublattices is superposed to highlight the sublattice polarization in the different regions. The dashed-line triangles indicate opposite orientations of the lattice polarization in each region.

**Figure 3. Simulated LDOS in buckled graphene. a-c**, Simulated LDOS evolution with the PMF amplitude for the crest, transition, and trough regions, respectively. Top (bottom) panels represent the A (B) sublattice. **d**, **Top panel:** STM topography ($V$ = 500 mV, $I$ = 20 pA) showing the triangular superlattice with the alternating crest (red) and trough (blue) regions. **Bottom panel:** PMF configuration used in the simulation. The symbols indicate the positions of the calculated LDOS in **a-c** and the green line indicates the calculated path for **g**. **e-f**, **Top panels:** Calculated LDOS as a function of energy in the crest and trough regions for $B_{eff}$ = 112 T ($B$ = 120 T). **Bottom panels:** Level-index dependence of the simulated (black square) and experimental (red dot) peak energy levels. The red line represents a fit to the experimental data.

$E_D = 0.43$ eV is taken from the experimental data on an unbuckled region of the sample. **g**, **Top panel:** Contour plot of the LDOS spectra (sublattice averaged) connecting two crests (along the green line in panel **d**). Green arrows indicate the positions of the peaks in the measured spectra shown in Fig. **2b** (right-pointing arrows) and Fig. **2c** (left-pointing arrows). **Bottom panels:** Evolution of the simulated value of $B_{PMF}/B$ along the green path in panel **d**.

**Figure 4. Flat-bands and LDOS maps. a,** Calculated band structured for a buckled graphene superlattice with period $a_b = 14$ nm and PMF amplitude $B_{eff} = 112$ T ($B = 120$ T). Inset: Superlattice mini-Brillouin zone, nested within the original Brillouin zone of flat graphene is shown together with the trajectory along which the band structure is calculated. **b, c,** Calculated LDOS contours (**b**) and measured $dI/dV$ maps (**c**) at $B_{eff} = 112$ T for the three energies ($E_0$-$E_D$, $E_1$-$E_D$, $E_2$-$E_D$) that correspond to the flat-band regions in panel **a** as indicated by the color coding.

**Figure 5. Flat-bands in buckled G/hBN. a,** STM topography of the buckled G/hBN sample ($V_b = -300$ mV, $I = 20$ pA). **b,** Height profile along the black line in **a** shows the ~3.5 Å height modulation of the buckling pattern. **c,** Gate voltage dependence of the spectra in the transition region shows the shift of the CNP (arrows) with doping ($V_b = -400$ mV, $I = 20$ pA). The black dashed line labels the Fermi level ($E_F$). **d,** $dI/dV$ spectra in the crest and transition regions labeled by the star and the circle in panel *a* respectively ($V_b = -400$ mV, $I = 20$ pA). Inset: PLL energy on crest area plotted against the square-root of the LL index, $N = 0, \pm 1, \pm 2, \ldots$ **e,** Gate voltage dependence of the spectra in the crest region shows the pseudo-gap feature and spectral weight redistribution in the partially filled flat-band ($V_b = -400$ mV, $I = 20$ pA).

# Methods

## Contents

0. Sample preparation
1. Buckling transition and pattern formation
2. STM results on flat (unstrained) G/NbSe$_2$ and G/hBN samples
3. Buckling patterns and dI/dV maps in the G/NbSe$_2$ sample
4. Transition region in the triangular buckling pattern
5. Dependence of the PMF on superlattice period
6. Effective PMF in the crest areas
7. Flat-band structure in buckled G/NbSe$_2$
8. Tight-binding model for the periodically strained triangular lattice
9. Theoretical results for G/hBN sample
10. Robustness of the PLLs against lattice disorder
11. Transition region in the rectangular buckling pattern of G/hBN

### 0. Sample preparation

To avoid oxygen and moisture contamination the heterostructures are fabricated in a dry Ar atmosphere in a glovebox. Graphene is first exfoliated on a PMMA film and then transferred onto a thin NbSe$_2$ flake deposited on an SiO$_2$/Si substrate. Instead of removing the PMMA, (poly methyl methacrylate) we mechanically peeled it off. The Au electrodes were deposited by standard SEM lithography. Before loading the sample into the STM chamber, the sample is annealed at 230 °C in forming gas (10% H$_2$ and 90% Ar) overnight to remove the PMMA residue. The STM experiments are performed in a home-built STM[27] at 4.6 K with a cut Pt$_{0.8}$Ir$_{0.2}$ tip and the sample are located using a capacitive-based self-navigation technique[36]. The tip used here is calibrated on the Au electrode, and the *dI/dV* curves are measured using a standard lock-in technique with a small A.C. voltage modulation (2 mV at 473.1 Hz) added to the D.C. junction bias[27].

### 1. Buckling transition and pattern formation

The periodic pattern formation following a buckling transition is largely determined by the boundary conditions and stress distribution[8,9]. In order to understand the buckling pattern in the G/NbSe$_2$ sample we carried out large area topography measurements that include the boundaries of the pattern (Extended Data Figure 1). In Extended Data Figure 1, A, the triangular pattern is delimited by two intersecting ridges (labeled as ridge 1 and ridge 2, ~0.5 μm long and 4 nm tall) that form a 60° fan. Zooming into the fan area (Extended Data Figure 1, C), we discern the buckling pattern. In Extended Data Figure 1, C and D, we show that the period of the pattern increases monotonically with the distance from the apex where the two ridges meet.

When graphene is deposited on a substrate, this process produces folds, ridges and bubbles due to trapped gas or solvents. Many of these defects disappear upon annealing (see Sample Preparation), but not all, presumably because their geometry is such that it does not allow the trapped species to escape or because the defect is pinned to the substrate. We observe that the ridges that survive the annealing step (Extended Data Figure 1, A) do not show the usual concave profile seen prior to annealing but rather show evidence of collapse, probably due to freezing of the trapped gas that supported the ridge. Following the collapse, the ridge profile becomes convex and is flanked by two tall lips on the boundaries of the original fold (Extended Data Figure 1, E). The lip pointing inwards towards the fan area is consistently shorter than the outside lip, suggesting that some of the graphene membrane comprising the original fold was pushed inwards. This increases the area of the membrane trapped between the two ridges (ridge 1 and 2 in Extended Data Figure 1 A) resulting in biaxial compressive strain which can trigger the buckling transition. To test this scenario, we conjecture that the concave region of the ridge was originally part of the convex top. This suggests that one can reconstruct the original shape by a mirror transformation of the concave part relative to the green dashed line in Extended Data Figure 1, E (the line intersects the tallest point of the convex part and is parallel to the substrate). This produces the reconstructed dome shown by the red symbols in Extended Data Figure 1, E. Using this procedure immediately reveals a missing part of the original dome of length $\Delta L$ as indicated in the figure. The strain produced by this increased length is estimated from the ratio between $\Delta L$ and the bisector of the 60° triangle: $\varepsilon = \Delta L/L \sin 30° = 2\Delta L/L$ where L is the distance from the apex formed by the intersection of the two ridges (Extended Data Figure 1, F).

Theoretical models and simulations of wrinkling based on minimizing the energy of a stretched or compressed membrane by allowing out of plane distortions[8,9], have shown that there are simple scaling laws relating the period of the buckled membrane $\lambda$ to the strain $\varepsilon$:

$$\frac{\lambda^4}{(tL)^2} = \frac{4\pi^2 \nu}{3(1-\nu^2)\varepsilon}.$$

Using t = 0.3 nm for the graphene thickness and $\nu \sim 0.15$ for its Poisson ratio and the expression for the strain as a function of distance L, we find

$$\lambda = \left(\frac{4\pi^2 vt^2}{3(1-v^2)440}\right)^{1/4} L^{3/4} = 0.14 L^{3/4}$$

Fitting the data for the distance dependence of the strain, shown in Extended Data Figure 1, D to the expression $\lambda = \lambda_0 + aL^{3/4}$ we obtain $a = 0.154 \pm 0.005$ consistent with the above estimate. The value of the offset $\lambda_0 = (6 \pm 0.3)\ nm$, suggests that this formula breaks down at short distances.

To understand why the 1D scaling of the buckling period is consistent with our results we consider the sketch shown in Extended Data Figure 1, B. First, we note that it is unlikely that both ridges collapse at the same instant. Now if we suppose that the ridge marked by the green line in the sketch collapses first, it will produce a set of roughly parallel wrinkles whose spacing increases with distance as the strain decreases according to the scaling formula above. When subsequently the yellow ridge collapses it produces a similar set of wrinkles roughly parallel to itself. The points of intersection of the two wrinkle sets will thus produce a triangular pattern of crests (black dots) while the areas in between will be troughs.

Buckling transitions in thin stiff membranes have been studied extensively both experimentally and theoretically in the context of mechanical engineering. The details of the patterns that emerge after the buckling transition has taken place, are controlled by boundary geometry and the strain distribution[8,9]. Depending on the value of these parameters a variety of buckling patterns are observed including square, hexagonal, herringbone, and stripes[7-9].

2. **STM results on flat (unstrained) G/NbSe$_2$ and G/hBN samples**

Superposing two 2D crystal structures produces a moiré pattern whose period is controlled by the two atomic lattice constants, $a$ and $b$, and by the angle between their crystal orientations. The largest moiré period, which is obtained when the two crystals are aligned, is given by $\lambda_{max} = (1+1/\delta)a$, where $\delta = (b-a)/a$, is the lattice mismatch[37]. The lattice constants for G and NbSe$_2$ are $a = 0.246$ nm and $b = 0.36$ nm, respectively, leading to $\lambda_{max} = 0.77$ nm. This is more than an order of magnitude smaller than the superlattice periods observed in our work, immediately ruling out an interpretation in terms of a moiré pattern.

To further confirm that the observed pattern is not due to a moiré structure we show in Extended Data Figure 2 the atomically resolved topography of a region far from the two ridges which shows the honeycomb structure characteristic of flat graphene (Extended Data Figure 2, B). In this region, the featureless STM topography (Extended Data Figure 2, A) together with the V-shaped *dI/dV* spectrum confirm that the graphene is well decoupled from the bottom NbSe$_2$. This procedure of taking spectra in an unbuckled region of the sample also served for tip selection in all our measurements. When the spectra outside of the buckled region showed an anomalous feature such as gap or large dip at the Fermi energy[38], the tip was reconditioned or replaced. The tip integrity is particularly important for distinguishing between correlation induced gap-features and artifacts introduced by the tip.

### 3. Buckling patterns and dI/dV maps in the G/NbSe2 sample

The two panels of Extended Data Figure 3, A show the STM topography of the buckled graphene taken with two different bias voltages. The blue lines are guides to the eye connecting the crests. The crests remain bright for different bias voltages consistent with their higher topography.

In Extended Data Figure 3, B we show the *dI/dV* map at an energy corresponding to the *N* = 0 pseudo Landau level in the crest area. The uniform LDOS represented by this map differs from the petal structure expected for a Gaussian bump[12,25] and reflects the unique geometry of the PMF induced by the buckled structure, which further confirms the theoretical model.

### 4. Transition area in the triangular buckling pattern

Extended Data Figure 4, A shows the theoretical contour plot of the LDOS spectra connecting two crest areas (also the upper panel of Figure 3g in the main text). In the transition area labeled by the yellow dashed line, we note that features from the crest (green arrows) and trough (red arrows) coexist. This coexistence is also observed in the experimental *dI/dV* spectrum (Extended Data Figure 4, B) which exhibits peaks that can be traced back to both the crest (green arrows) and the trough (red arrows) regions.

### 5. Dependence of the pseudo magnetic field (PMF) on superlattice period

Extended Data Figure 5 shows the simulated LDOS in the crest region as a function of PMF amplitude for several values of the super-period, $a_b$ , as marked. The dashed lines represent the field-dependence of the Landau level energy in uniform fields. We note that as the lattice spacing increases the spectra start approaching the unstrained LL sequence at lower values of the PMF. This is consistent with the fact that LLs, which correspond to cyclotron motion, can only form if the magnetic field is fairly constant over length scales that are several times the magnetic length, $l_B = \sqrt{\frac{\hbar}{eB}} \approx \frac{25.7nm}{\sqrt{B}}$.

6. **Effective PMF ($B_{eff}$) in the crest area**

Extended Data Figure 6, A shows LDOS cuts in the crest region from Figure 3a in the main text (A sublattice) for several values of B, as marked. For each spectrum we calculated the value of the effective magnetic field, $B_{eff}$ from the energy of the first peak $E_1 = v_F\sqrt{2e\hbar B_{eff}}$. In order to check if the peaks can be interpreted as LLs we mark by arrows the calculated peak sequence for the corresponding $B_{eff}$, $E_N = \pm v_F\sqrt{2e\hbar B_{eff} N}$. We note that the arrows coincide with the spectral peaks for sufficiently large PMF amplitudes suggesting that the Landau level language is appropriate for these spectra, as long as the $B_{eff}$ is used. However, the extracted effective field is much smaller than the maximum PMF value, *i.e.* $B_{PMF}^{max} = 3B$, reflecting the fact that the cyclotron orbit averages the field over an area of size $\sqrt{2N+1}l_B$. For example, for a 14nm lattice period and $B = 120$ T the peak sequence gives $B_{eff} = 112$ T, which is consistent with the experimental result in Figure 2b.

As shown in Figure 3a and Extended Data Figure 5 the *N*=0 PLL undergoes a transition from a double peak at low PMF to a single one at a higher PMF at a critical PMF value which is highly sensitive to the lattice spacing. This is clearly illustrated in Extended Data Figure 6, B where we show that the simulated *N*=0 double-peak for $B_{eff} = 112$ T and $a_b = 14$ nm, merges into a single peak for $a_B = 14.8$ nm. The data in Figure 2 of the main text, where: $a_B = 14.4$ nm $\pm 0.5$ nm, $B_{eff} = 112$ T and the *N*=0 PLL peak is unsplit - is consistent with the simulated spectrum (Extended Data Figure 6, B red curve) in the regime above the critical PMF.

7. **Flat-band structure in buckled G/NbSe$_2$**

In Extended Data Figure 7 we plot the band structure and LDOS in the trough region for $B = 140$ T (Extended Data Figure 7A, 7B) and $B = 180$ T (Extended Data Figure 7C, 7D). The figure shows the flattening of the bands with increasing field amplitude.

8. **Tight-binding model for the periodically strained triangular lattice**

The effect of strain is included in the tight-binding Hamiltonian through the modulation of the hopping energy. This is given by

$$t_{ij} = t_0 e^{-\beta(d_{ij}/a_{cc}-1)} \qquad (1)$$

where $d_{ij}$ is the strained bond length defined by the strain tensor $\bar{\varepsilon}$ as

$$d_{ij} = (I + \bar{\varepsilon})\delta_{ij}, \qquad (2)$$

where $\delta_{ij}$ is the vector in the direction of the bond between atoms $i$ and $j$ and $I$ is the unitary matrix. Changes in the hopping energies generates a strain induced vector potential in the system which in the case of hexagonal lattices is given by

$$A_x - iA_y = -\frac{1}{ev_F}\sum_j \delta t_{ij} e^{i\mathbf{K}\cdot\mathbf{r}_{ij}}. \qquad (3)$$

The corresponding PMF is calculated using

$$\mathbf{B} = \nabla \times \mathbf{A} = [\partial_x A_y - \partial_y A_x]\mathbf{e}_z. \qquad (4)$$

Due to gauge invariance we may choose $A_y = 0$. Hence, our vector potential is then given by $A_x = \int B(x,y)dy$, where $B(x,y)$, shown in Extended Data Figure 8, A, is given by Eq. (1) of the main manuscript. This leads to

$$A_x = B_0 \frac{a_B}{2\pi}\left[\frac{1}{b_{1y}}\sin(\mathbf{b}_1 \mathbf{r}) + \frac{1}{b_{2y}}\sin(\mathbf{b}_2 \mathbf{r}) + + \frac{1}{b_{3y}}\sin(\mathbf{b}_3 \mathbf{r})\right]. \qquad (5)$$

Substituting $t_{ij} = t_0(1 + \delta t_{ij})$ and expanding Eq. (3) up to first order, we obtain the following expression for the vector potential[6]

$$(A_x, A_y) = -\frac{1}{2ev_F}\left(2\delta t_1 - \delta t_2 - \delta t_3, \frac{1}{\sqrt{3}}(\delta t_2 - \delta t_3)\right), \qquad (6)$$

where $\delta t_1$, $\delta t_2$, and $\delta t_3$ are the strain modulations of hopping energies along the directions of graphene's nearest neighbors $\delta_1$, $\delta_2$, and $\delta_3$, as shown in Extended Data Figure 8, B, and $v_F = 3t_0 a_{cc}/(2\hbar)$ is the Fermi velocity. The choice of the gauge ($A_y = 0$) results into $\delta t_2 = \delta t_3 = \delta t$. We choose $\delta t_1 = -\delta t$ and, finally, the strain modified hopping energies are given by

$$t_1 = t_0\left(1 - \frac{3A_x \pi a_{cc}}{2\phi_0}\right) \qquad (7)$$

$$t_2 = t_3 = t_0 \left(1 + \frac{3A_x \pi a_{cc}}{2\phi_0}\right),$$

where $\phi_0 = h/e$ is the magnetic flux quantum.

## 9. Theoretical results for G/hBN sample

To simulate the buckling pattern in graphene on hBN we use a similar expression as in the case of graphene on NbSe$_2$, given by Eq. (1), but without the last cosine term:

$$\boldsymbol{B_{PMF}}(x, y) = B[\cos(\boldsymbol{b_1}\boldsymbol{r}) + \cos(\boldsymbol{b_2}\boldsymbol{r})], \tag{8}$$

with, $\boldsymbol{b_1} = \frac{2\pi}{a_B^x}(0, 1)$, $\boldsymbol{b_2} = \frac{2\pi}{a_B^y}(1, 0)$. This changes the symmetry of the unit cell from triangular to rectangular. The profile of the field is shown in Extended Data Figure 9, A with the unit cell marked by the dashed rectangle. The size of the unit cell is chosen so as to match the buckling periods shown in Figure 5a. The LDOS in the crest region is given in Extended Data Figure 9, B where clear Landau levels can be observed in the spectra. A cut of the LDOS map at a constant value of $B = 62$ T (white dashed line in Extended Data Figure 9, B) shows LLs that scale with the square-root of the LL index (see Extended Data Figure 9, E and G) and result in an effective field of 100 T. Moving towards the transition region (magenta square in panel A) where $B = 0$ T, the LDOS spectrum changes significantly, as shown by Extended Data Figure 9, C. Here, the spectrum does not show LLs, which is expected since there is no PMF. However, the LDOS reveals new sets of peaks that weakly depend on the amplitude of the field. This is further confirmed in Extended Data Figure 9, D which shows that the dispersion of these peaks with unit cell size $a_b$ is consistent with that expected of confinement states. Taking a cut of the LDOS map from Extended Data Figure 9, C, at $B = 62$ T results in equidistant peaks separated by ~83 meV as shown in Extended Data Figure 9, F and H.

## 10. Robustness of the PLLs against lattice disorder

As seen in Figure 1C, the rectangular buckling array in the G/hBN sample is not perfectly periodic as implied by the double cosine potential used in the simulation. It is then natural to ask how robust is the simulated DOS and corresponding *dI/dV* spectrum against lattice distortions. To address this question we carried out numerical simulations where the periodicity condition was relaxed. As detailed below, we find that the PMF spectrum of the rectangular buckling structure is quite robust, and can survive significant deviations from the perfectly periodic lattice structure.

To study the effect of relaxing the condition of perfect periodicity we introduced a random variation in the unit cell of the PMF within a predefined range. The procedure is as follows. Firstly, since the original unit cell is too large, we reduced the larger unit vector to the size of the smaller one, i.e. $a_x = a_y = 20$ $nm$. The PMF profile for this case is shown in Extended Data Figure 10, A. Changing the size of the unit cell shouldn't change the physics of the problem, however numerically this significantly speeds up our calculations. The system is a circle of radius of 200 nm with absorbing boundary conditions applied at the edges [the method is described in detail in reference [39]. A vector potential is included into the Hamiltonian as described above with the unit cell period given by [$(1 + \Delta R_x)a_x$, $(1 + \Delta R_y)a_y$ ] where $R_x$ and $R_y$ are random numbers from uniform distribution in range [-1, 1] and $\Delta$ is the weight parameter. As an example, in Extended Data Figure 10, (B-D) we plot a few PMF profiles using $\Delta = 0.15$, 0.25, 0.33, respectively. The calculated DOS is shown in Extended Data Figure 10, E [DOS is calculated using the kernel polynomial method as explained in reference[40]. The blue curve in this plot shows the DOS of the perfectly periodic system. The peaks observed here are the flat-bands, and the separation between the peaks is around 60 meV. As $\Delta$ is increased the peaks are averaged out and eventually disappear for large $\Delta$ which is the expected result. However, up to $\Delta = 0.25$, the peaks are still present in spite of the significant disorder introduced in the system.

In addition to the robustness of the PLL spectrum against lattice disorder, the resemblance between the spectral features of the G/hBN and G/NbSe$_2$ samples in the crest regions, further indicates their common origin, as detailed below.

1) In both systems, the features are observed only in the buckled regions of the graphene membrane and are absent in the unbuckled parts of the sample.

2) In both systems the buckling gives rise to the strain-induced PMF resulting in the PLL observed in the *dI/dV* spectra. The PLL sequence, which is characterized by a peak at the Dirac point and a square-root dependence on level index, is observed on the crests, consistent with the numerical simulation results.

3) In both systems the spectra in the regions between maxima of the PMF show confinement character, rather than PLLs. Specifically, they do not feature the peak at the Dirac point corresponding to the *N*=0 PLL and the peak sequence does not follow a square-root level index dependence. Instead the spectrum in this regime is consistent with magnetic confinement states.

## 11. Transition region in the rectangular buckling pattern of G/hBN

In the case of the G/NbSe$_2$ sample, the crest and trough regions of the triangular buckling pattern observed in topography coincide with the maximum and minimum PMF respectively. This designation is not as straightforward in the rectangular buckling pattern observed in the G/hBN sample. While in this sample the crest regions where the PMF has its maximum magnitude are readily identified in the topography as the intersection point between two of the quasi 1D wrinkles that form the rectangular buckling pattern (Figure 1c in the main text), identifying the transition regions where the PMF is vanishingly small, is less obvious. We instead used the fact that in the transition region the PMF vanishes resulting in the disappearance of the PLLs and its characteristic peak at the CNP. By using this criterion, we identified the transition region marked by the circle in Figure 5a of the main text.

**Method References:**

**Data Availability**
The data that support the findings of this study are available from the corresponding authors on reasonable request.

**Extended Data Figure Legends**

**Extended Data Figure 1. A,** Large area STM topography of G/NbSe$_2$ shows two ridges that delimit the buckling pattern ($V_b$ = -0.3 V, I = 20 pA). **B**, Schematics of wrinkles arising from the compressive strain at each boundary ridge. Crests form at the wrinkle intersections are marked by black circles. **C**, Zoom-in topography image of the triangular buckling pattern ($V_b$ = 0.5 V, I = 20 pA) in A. **D**, The period of the buckling pattern, measured along the line marked by the blue arrow in C, increases monotonically with distance from the apex where the two ridges meet. **E**, Height profile along the green arrow in A. **F,** Strain produced by the collapse of the ridges calculated from E, as described in the text (Methods 1).

**Extended Data Figure 2. A**, STM topography in a flat (unbuckled) region of the G/NbSe$_2$ surface far from the ridges. Inset: $dI/dV$ spectrum on (A). $V_b$ = 0.5 V, I = 30 pA. **B**, Atomic resolution of G/NbSe$_2$ in (A). $V_b$ = -0.3 V, I = 30 pA. **C,** Same as (A) but in a flat region of the G/hBN sample. Inset: $dI/dV$ spectra on flat G/hBN for electron-doping (green) and hole-doping (red). $V_b$ = -0.3 V, I = 20 pA.

**Extended Data Figure 3. A**, STM topography of a region in the buckled graphene membrane measured with different bias voltages, 500 mV (**Left**) and 50 mV (**Right**). **B**, $dI/dV$ map over an area of size 6×6 nm$^2$ in the crest region at the energy of the $N = 0$ PLL.

**Extended Data Figure 4. A**, Theoretical contour plot of the LDOS spectra connecting two crest areas (upper panel of Figure 3g in main text). Yellow dashed line labels the LDOS cut along the transition position. **B,** Experimental $dI/dV$ spectrum in the transition region between crests and troughs. Green and red arrows indicate the corresponding peaks in (A).

**Extended Data Figure 5.** Evolution of the LDOS with the PMF amplitude calculated for several supperlattice periods. Dashed lines represent the field-dependence of the PLL energy in uniform fields.

**Extended Data Figure 6. A,** LDOS cuts in the crest region from Figure3a (top panel) in the main text (A sublattice) for several values of B and corresponding $B_{eff}$. The ratio of the effective PMF obtained from the PLL spectrum on the crests to the maximum PMF value is indicated in the far right legend. **B,** Comparison of crest LDOS at $B_{eff}$ = 112 T for two lattice constants (14.0 nm and 14.8 nm) in the regime where the $N$=0 peaks merge into one.

**Extended Data Figure 7.** Low energy band structure and LDOS in the trough region for two values of the PMF, 140 T (**A-B**) and 180 T (**C-D**), calculated for a superlattice period of 14 nm.

**Extended Data Figure 8. A**, Calculated profile of the PMF given by Eq. (1) in the main manuscript for lattice spacing $a_b = 14$ nm. **B**, Schematics of nearest neighbor vectors, $\delta_1$, $\delta_2$, and $\delta_3$, in a graphene lattice.

**Extended Data Figure 9. A,** Profile of the PMF in a rectangular buckling pattern. Dashed rectangle shows the magnetic unit cell. **B**, LDOS (sublattice averaged) versus the PMF

amplitude and energy at a crest position in A (center of red region marked by green dot). The color bar represents the strength of the LDOS in arbitrary units. Note that the same result is obtained in the center of troughs (blue regions) where the PMF sign is reversed. **C,** The same as (B) but for a point with zero field, marked in (A) by the magenta square. **D,** Evolution of confinement levels with the size of the unit cell. **E,** Cut of the LDOS from (B) for a constant value of the field, $B = 62$ T, shown by dashed white line. **F,** Cut of the LDOS from (C) for a constant value of the field, $B = 62$ T, shown by the dashed white line. **G,** Fitting the peak sequence to a square-root dependence on the PLL index, gives an effective PMF of 100 T. **H,** The peak sequence in (F) gives a linear dependence on peak index with an energy spacing of ~83 meV.

**Extended Data Figure 10.** DOS versus unit cell size in the presence of lattice disorder. (**A-D**) The PMF profile used for calculating the DOS in panel E. The unit cell period variation range is given by $(1 \pm \Delta)\, a_0$ with $a_0 = 20$ nm and $\Delta = 0, 0.15, 0.25$, and $0.33$, respectively. (**E**) DOS obtained for different values of $\Delta$ given in the legend for $B = 62$ T (as in Extended Data Figure 9).

Figure 1

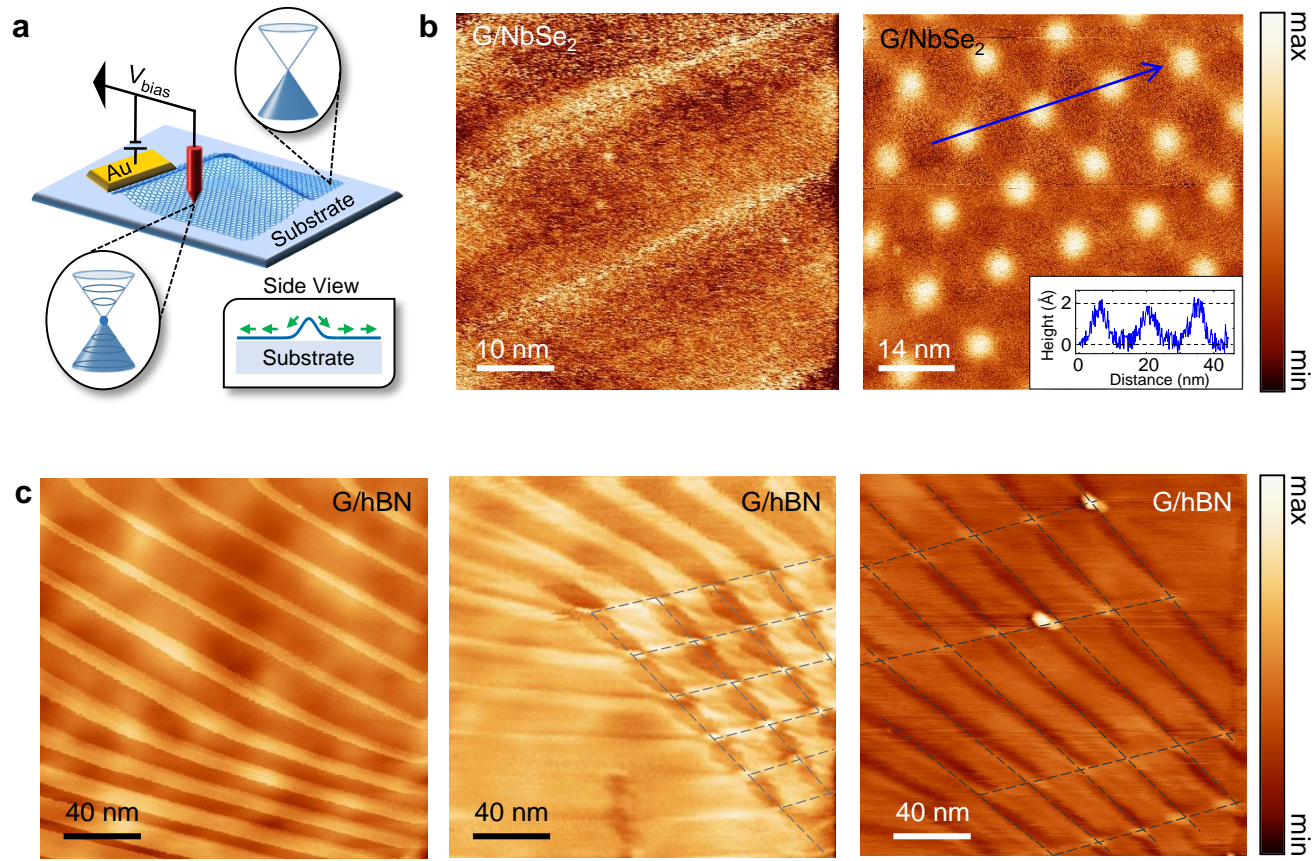

Figure 2

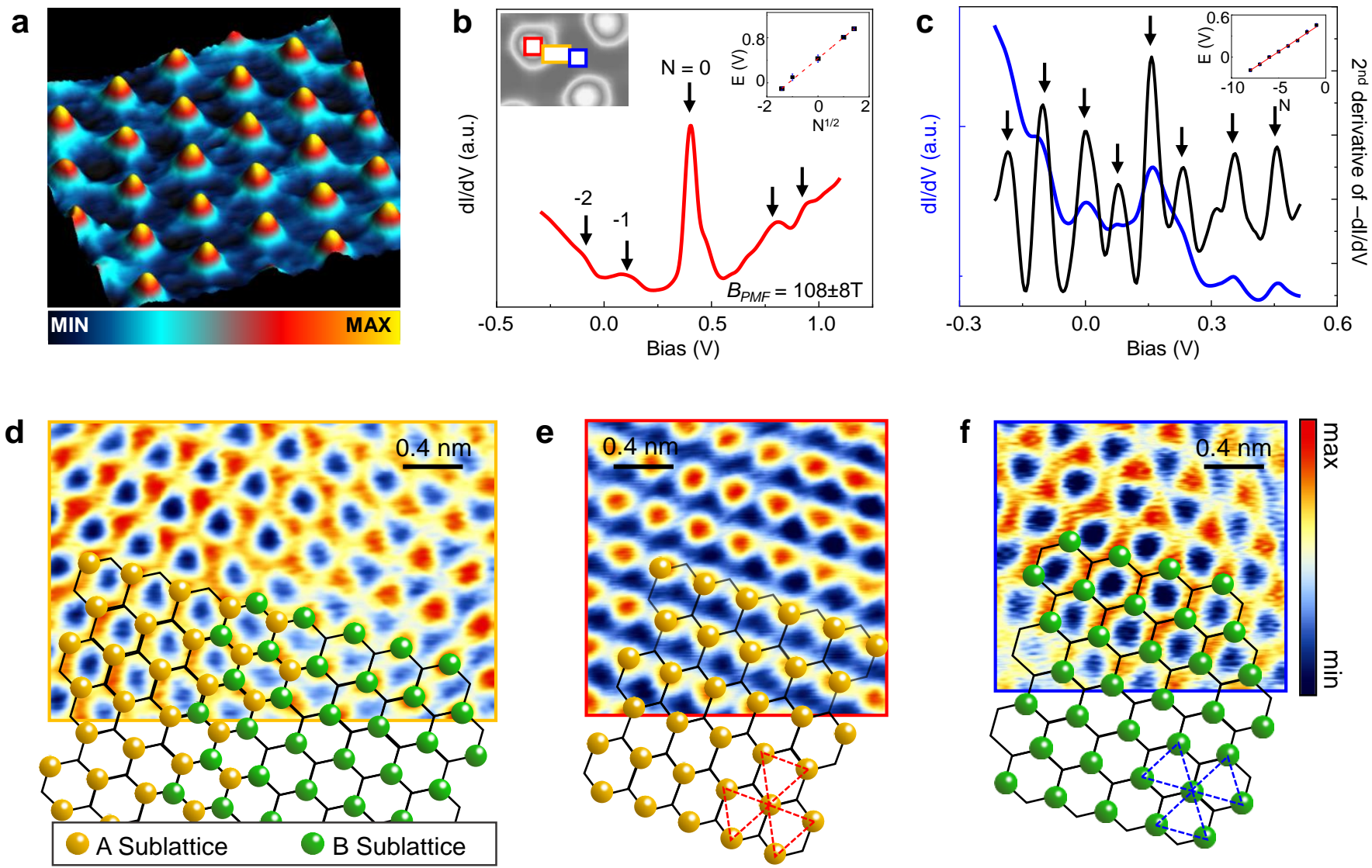

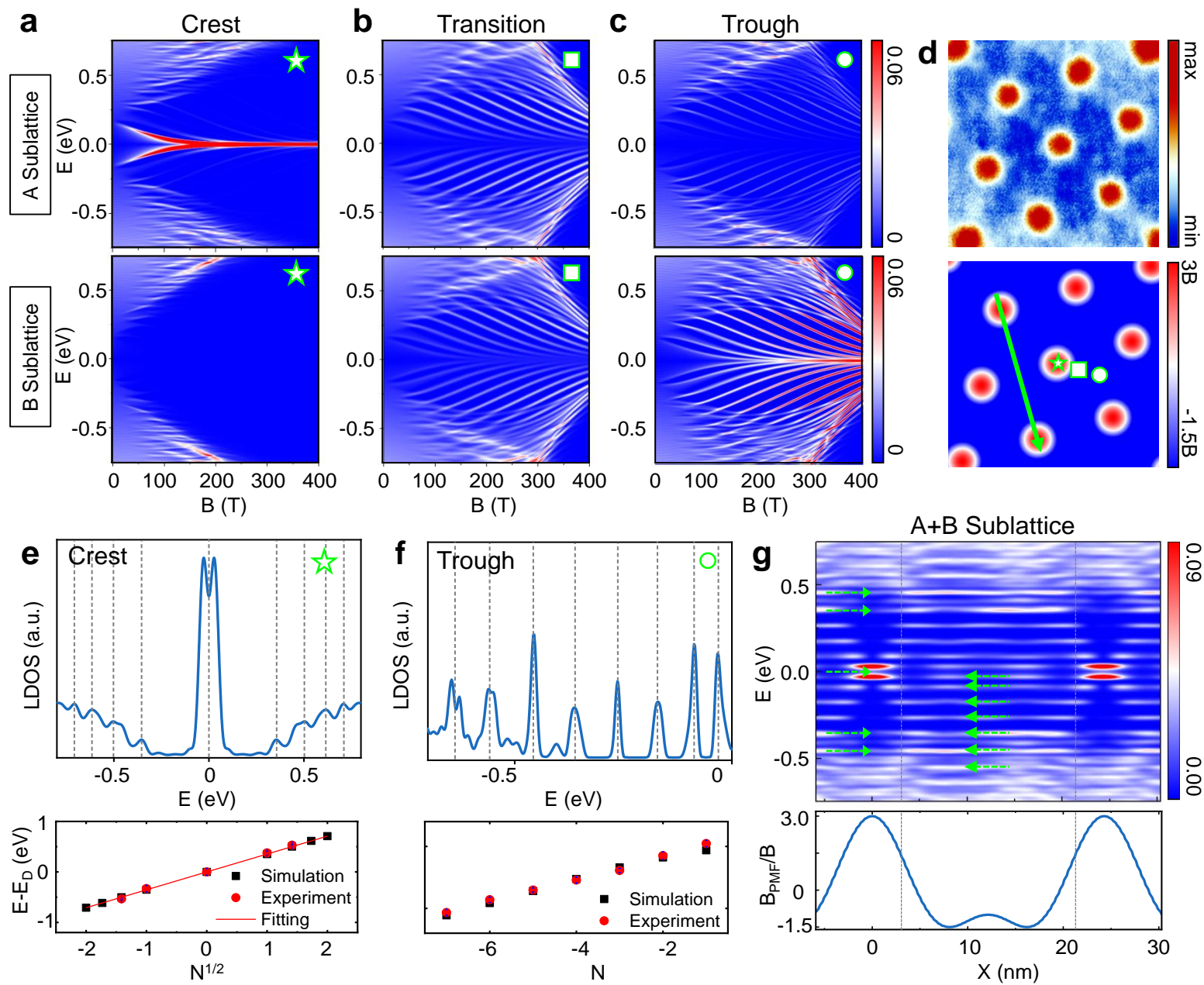

Figure 4

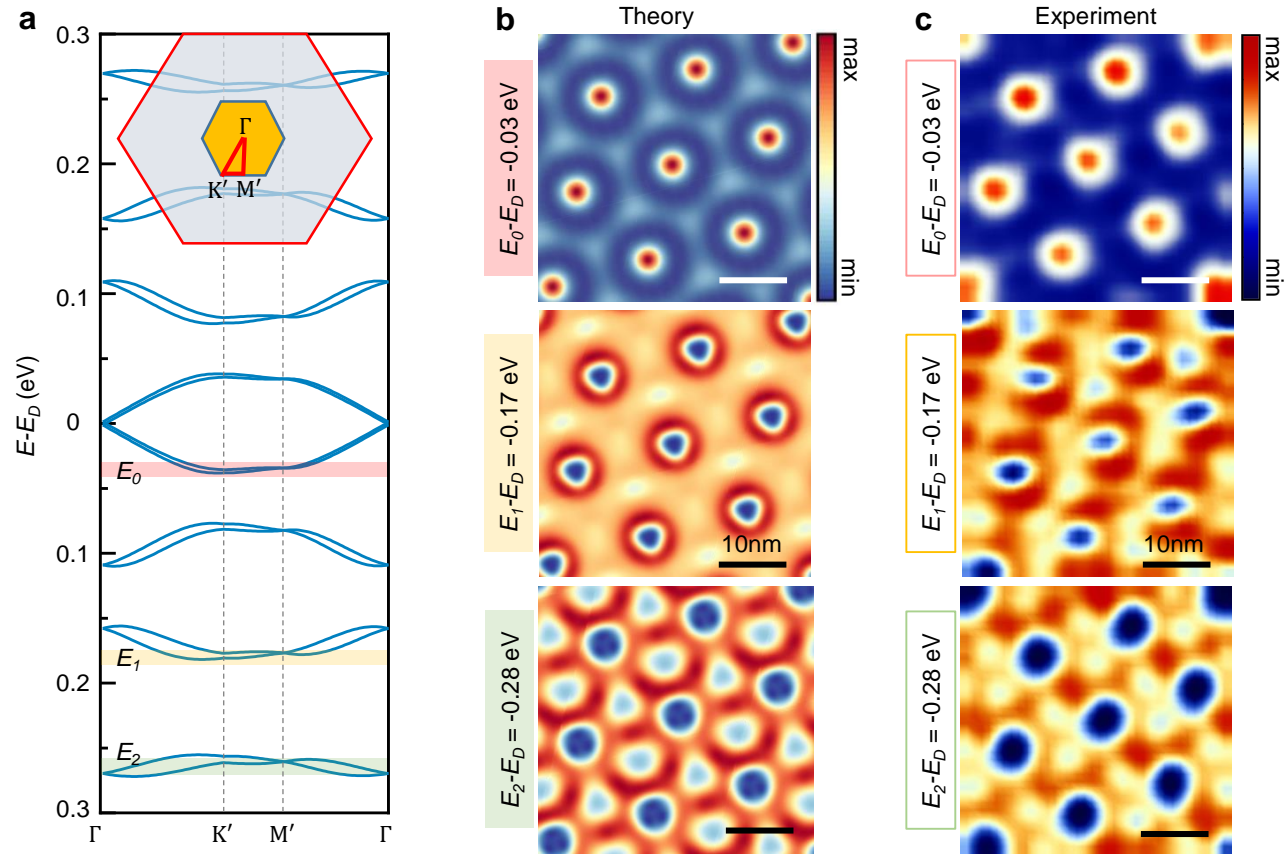

Figure 5

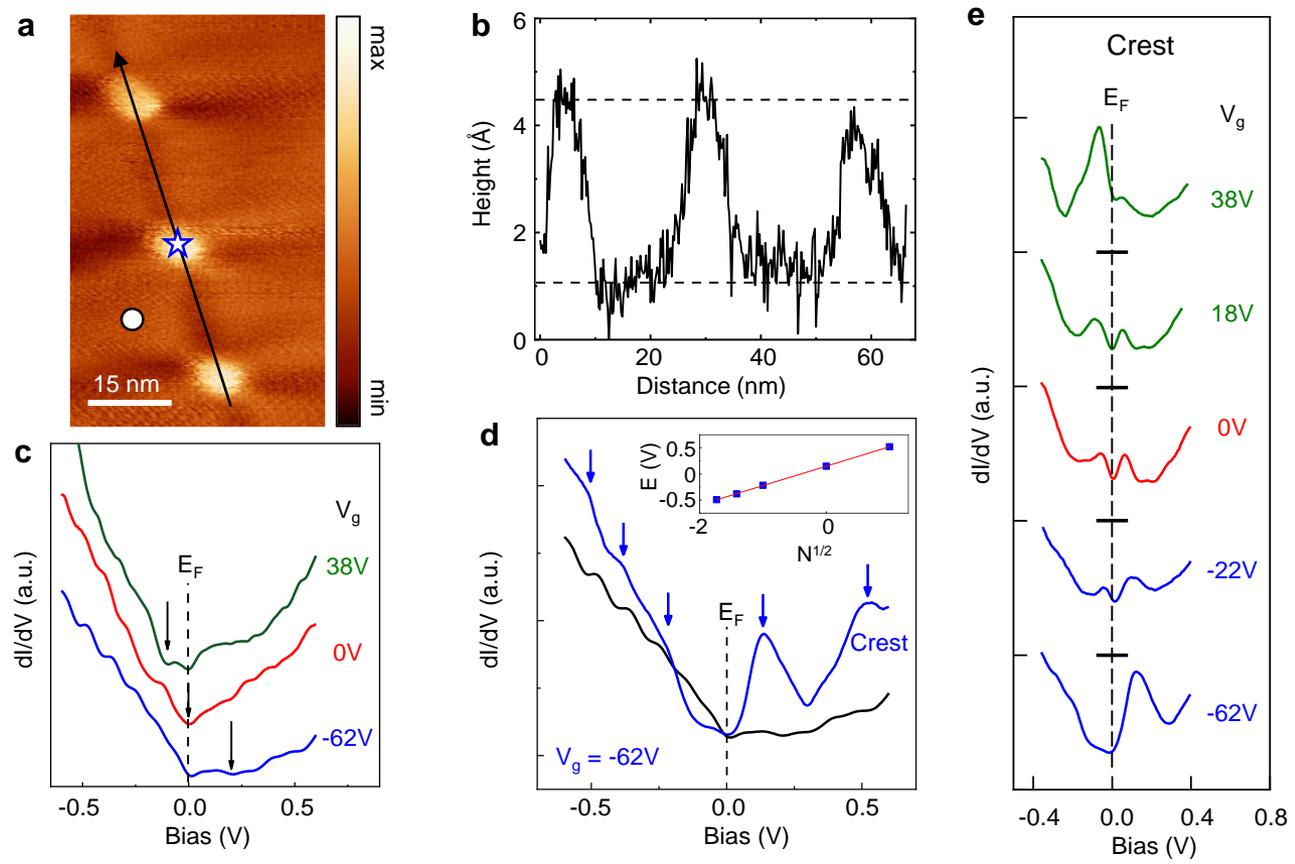

Extended Data Figure 1

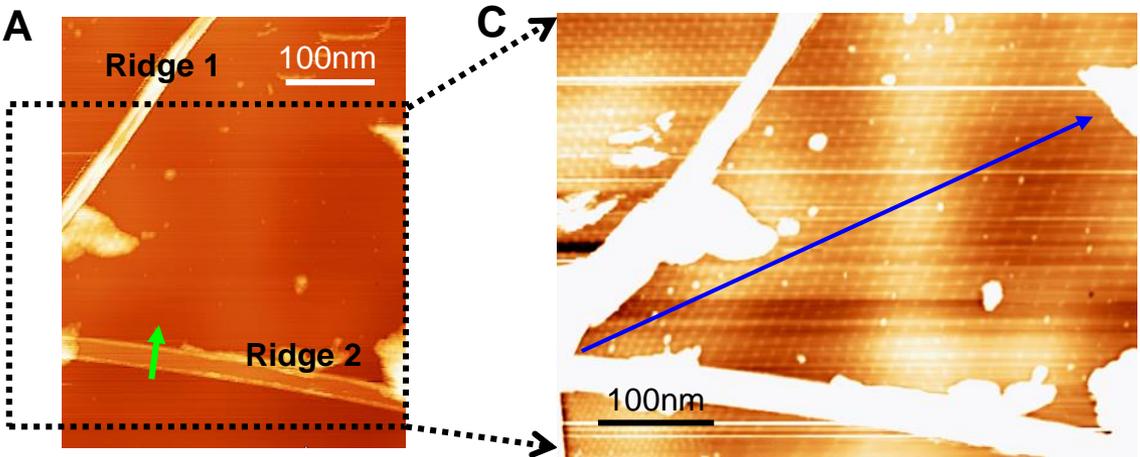
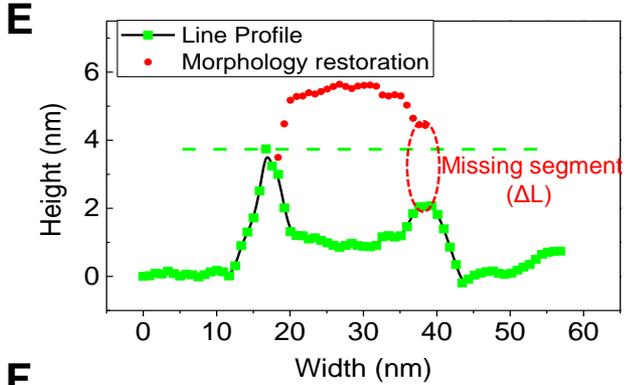
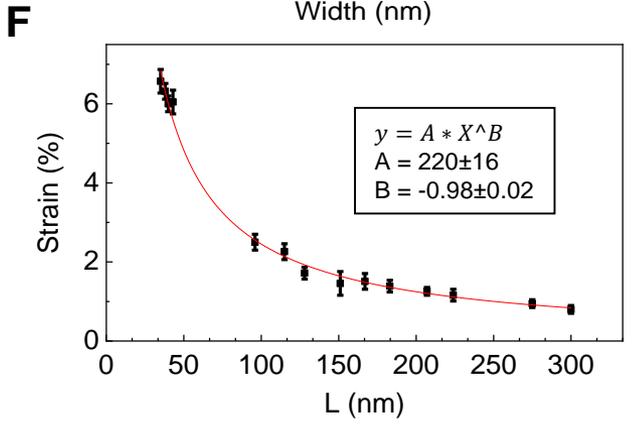

Extended Data Figure 2

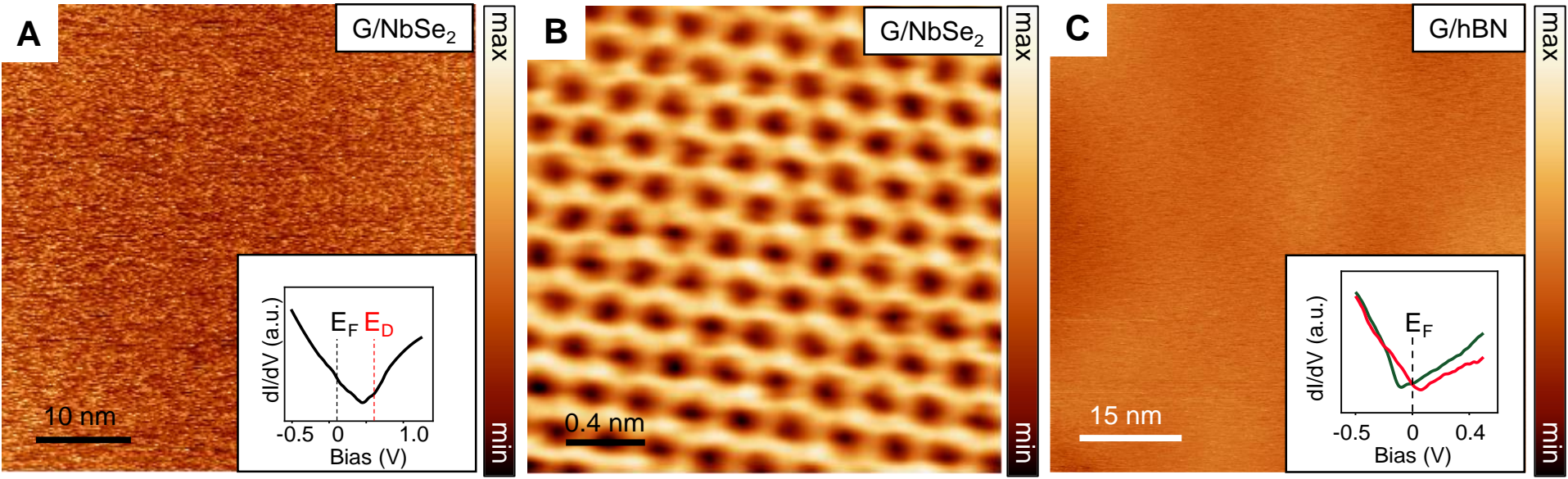

Extended Data Figure 3

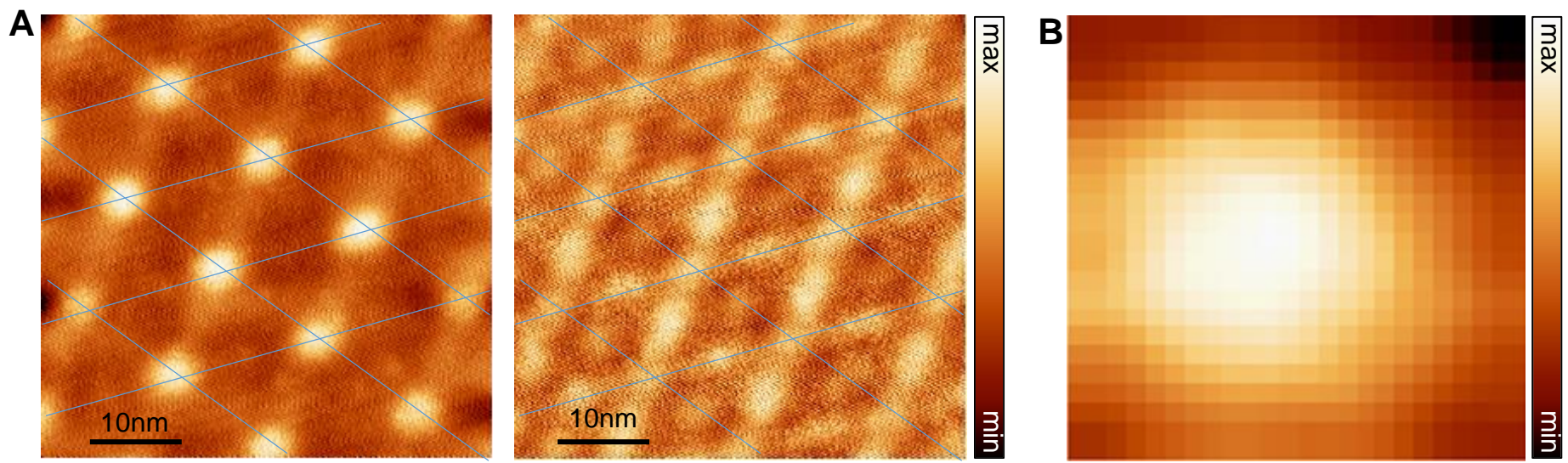

Extended Data Figure 4

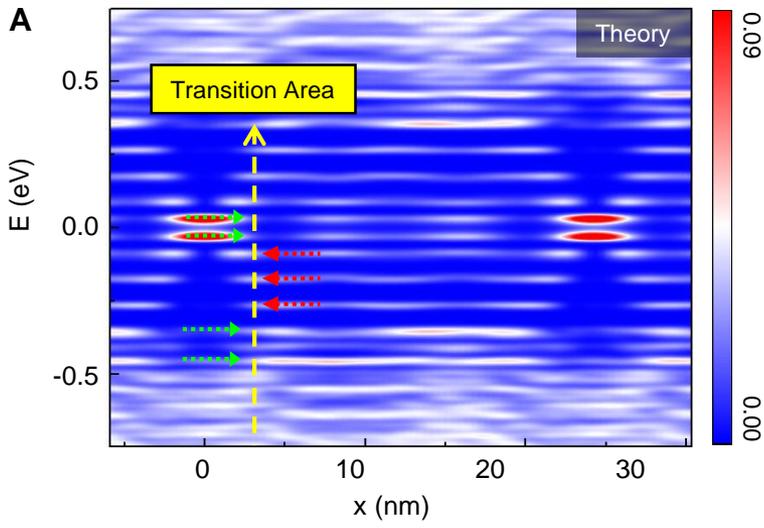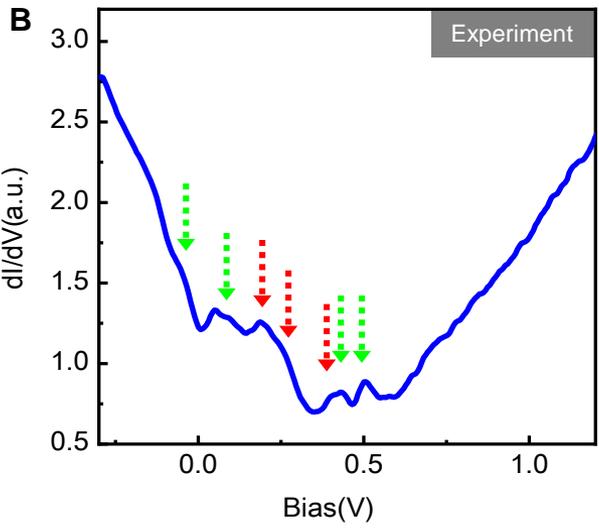

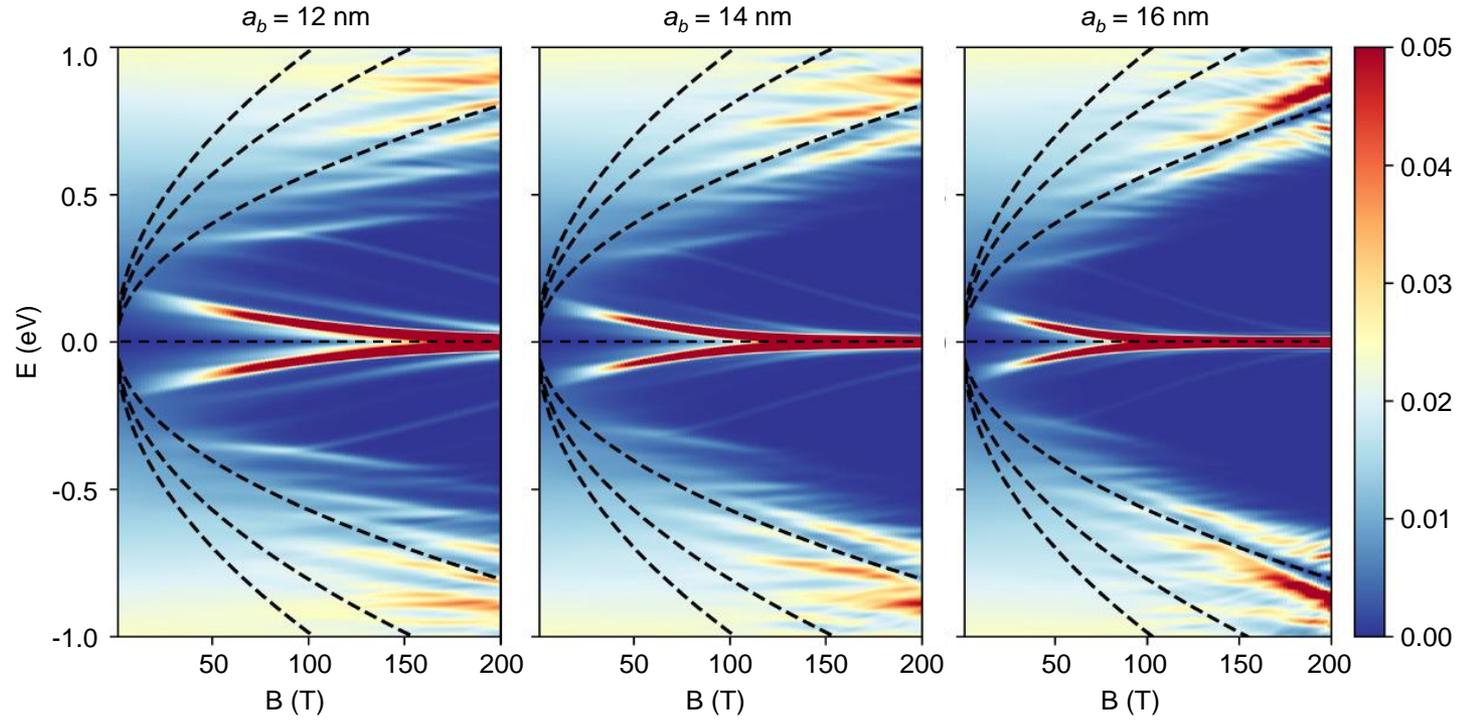

Extended Data Figure 6

**A**

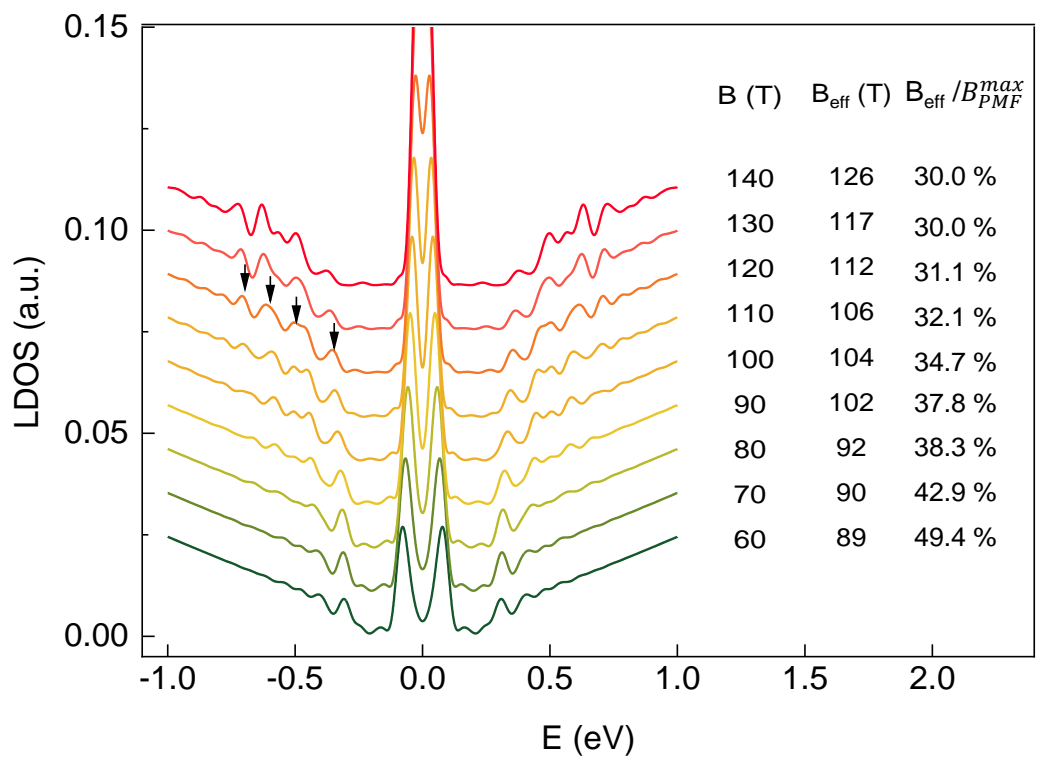

| B (T) | $B_{eff}$ (T) | $B_{eff}/B_{PMF}^{max}$ |
|---|---|---|
| 140 | 126 | 30.0 % |
| 130 | 117 | 30.0 % |
| 120 | 112 | 31.1 % |
| 110 | 106 | 32.1 % |
| 100 | 104 | 34.7 % |
| 90 | 102 | 37.8 % |
| 80 | 92 | 38.3 % |
| 70 | 90 | 42.9 % |
| 60 | 89 | 49.4 % |

**B**

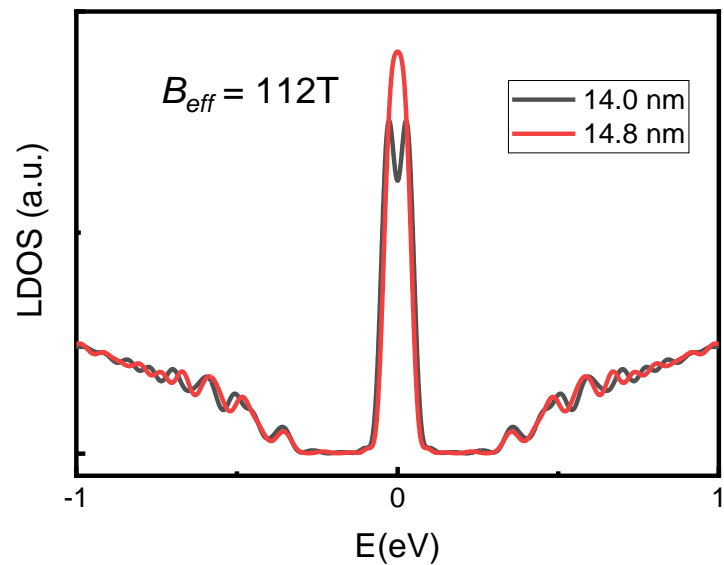

$B_{eff}$ = 112T

Extended Data Figure 7

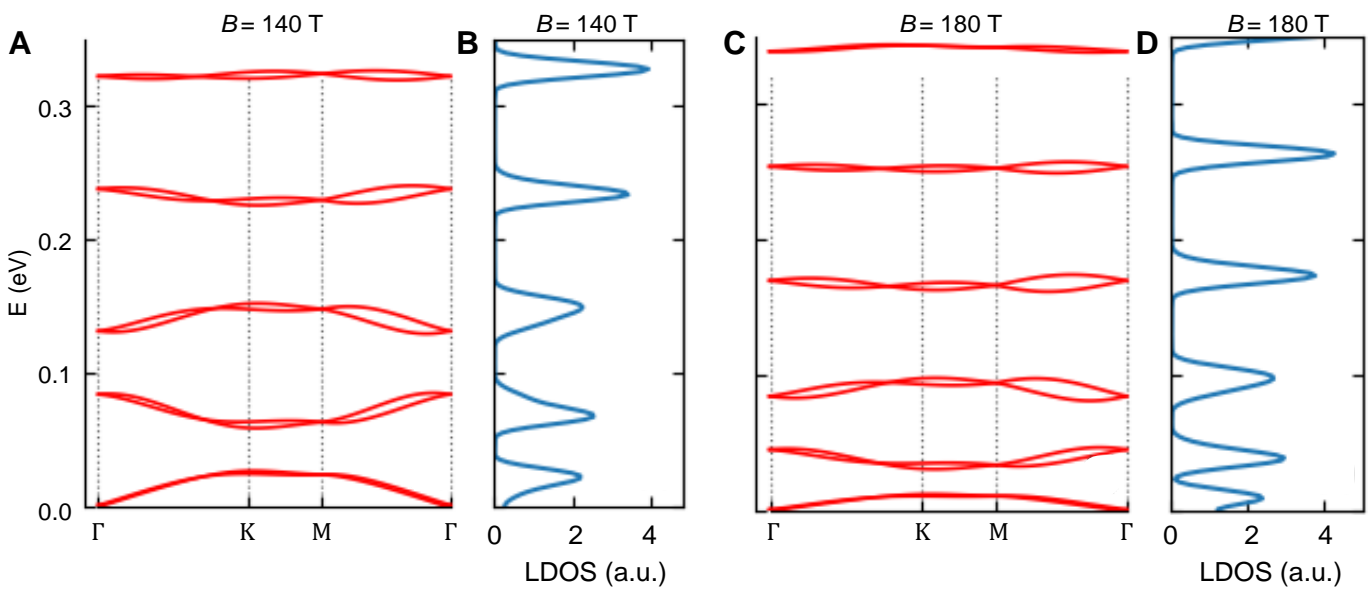

Extended Data Figure 8

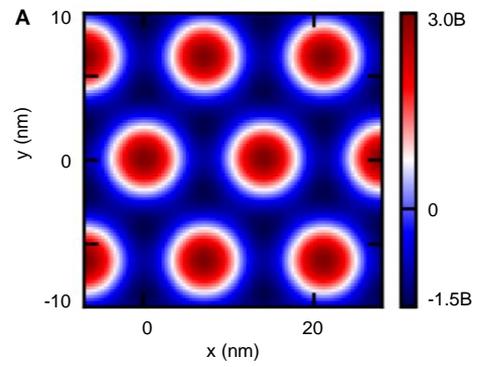 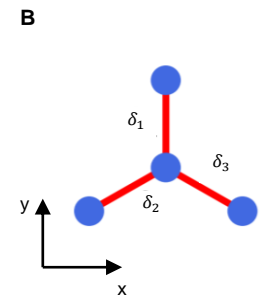

Extended Data Figure 9

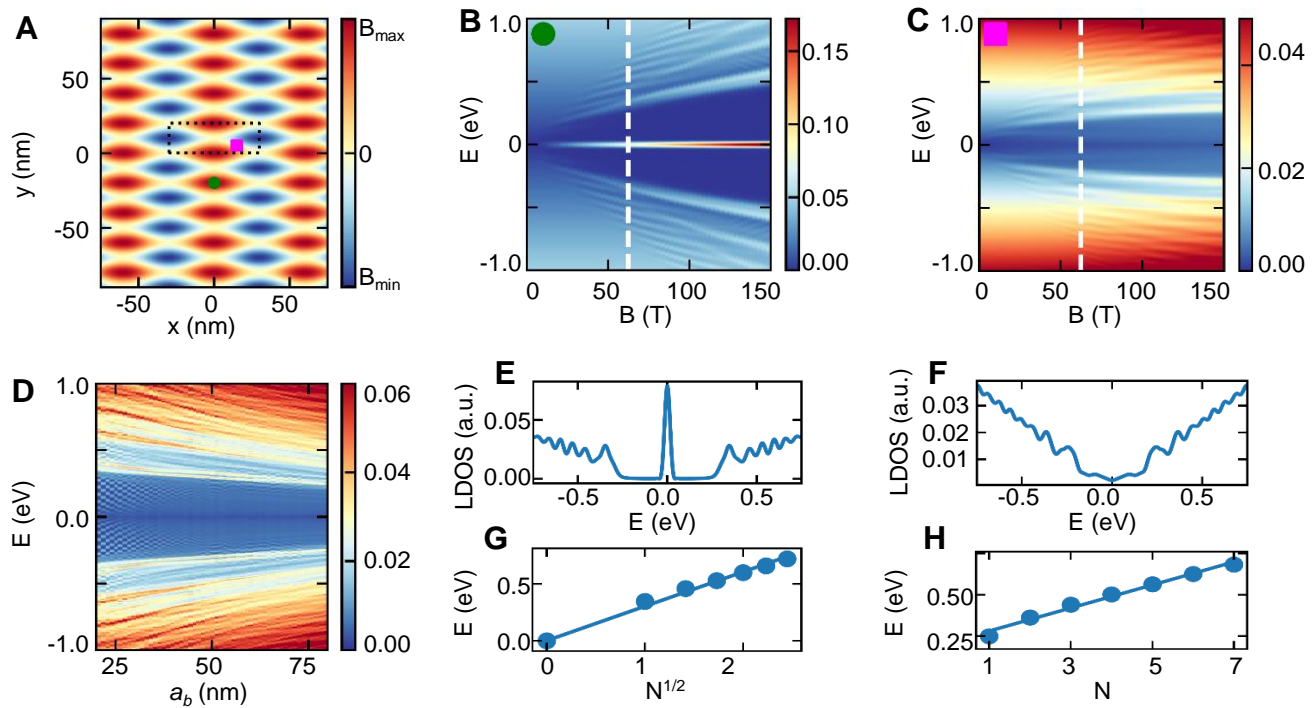

Extended Data Figure 10

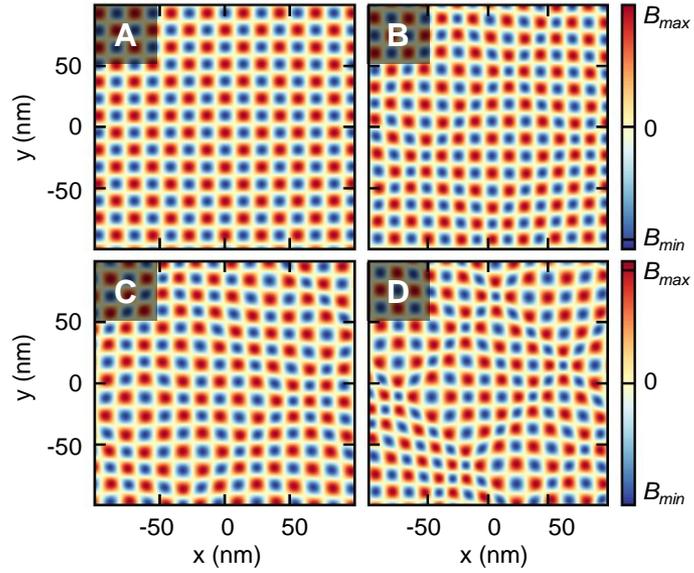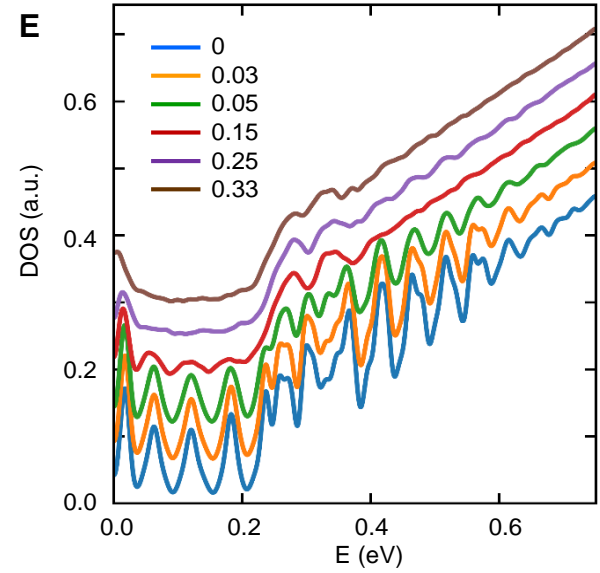